\documentclass[twocolumn]{aastex62}

\usepackage{apjfonts}

\shorttitle{Planetary Nebula in Open Cluster AL 1}
\shortauthors{Bellini et al.}

\newcommand{\Ha}{H$\alpha$}
\newcommand{\Hb}{H$\beta$}

\newcommand{\masyr}{\rm mas\, yr^{-1}}

\def\caii{\ion{Ca}{2}}

\def\nii{\ion{N}{2}}

\def\oiii{\ion{O}{3}}

\def\Gaia{{\it Gaia}}

\newcommand{\HST}{{\it HST}}

\begin{document}


\title{Testing Cluster Membership of Planetary Nebulae with High-Precision Proper Motions. II\null. \\ \HST\/ Observations of PHR J1315$-$6555 in the Open Cluster AL~1 (ESO 96-SC04) }

\author[0000-0003-3858-637X]{Andrea Bellini}
\affil{Space Telescope Science Institute, 
3700 San Martin Dr.,
Baltimore, MD 21218, USA}

\author[0000-0003-1377-7145]{Howard E. Bond}
\affil{Department of Astronomy \& Astrophysics, Penn State University, University Park, PA 16802, USA}
\affil{Space Telescope Science Institute, 
3700 San Martin Dr.,
Baltimore, MD 21218, USA}

\author[0000-0001-6008-1955]{Kailash C. Sahu}
\affil{Eureka Scientific Inc., 2542 Delmar Ave., Suite 100, Oakland, CA 94602, USA}
\affil{Space Telescope Science Institute, 
3700 San Martin Dr.,
Baltimore, MD 21218, USA}

\correspondingauthor{Howard E. Bond}
\email{heb11@psu.edu}

\begin{abstract}

Planetary nebulae (PNe) shown to be members of star clusters provide information on their properties and evolutionary histories that cannot be determined for PNe in the field, in particular the initial masses of their progenitor stars. Here we investigate the bipolar PN PHR\,J1315$-$6555 (hereafter PHR\,J1315), which lies near the open cluster AL\,1 (ESO 96-SC04) on the sky. Previous work has established that the PN and cluster have similar radial velocities and amounts of interstellar reddening, and similar distances estimated using independent methods. We have obtained new images of the PN and cluster using the {\it Hubble Space Telescope\/} (\HST). Combined with archival \HST\/ frames taken 12 years earlier, they provide high-precision proper motions (PMs) for two candidate central stars of PHR\,J1315. We find that the PMs of both candidates are consistent with those of cluster members, strongly confirming the PN's membership in AL\,1. The candidate lying closer to the center of PHR\,J1315 has the color and luminosity of an early F-type dwarf, suggesting that it may be the optical primary in a close post-common-envelope binary. We used the \HST\/ data to construct a color-magnitude diagram for AL\,1, which we corrected for significant foreground differential reddening. Isochrone fitting reveals that the cluster lies at a remarkably large distance of about 13~kpc, and has an age of about 1.0~Gyr. The initial mass of the progenitor of PHR\,J1315 was about $2.1\,M_\odot$. We suggest followup investigations that would provide tighter constraints on the object's evolution.
 
\null\vskip 0.2in

\end{abstract}



\section{Introduction: Planetary Nebulae in Star Clusters \label{sec:intro} }

If a planetary nebula (PN) is verified to be a member of a star cluster, it becomes possible to derive constraints on the mass and composition of the PN's progenitor star---information that cannot be obtained for PNe in the field. Moreover, the absolute luminosity of the PN nucleus (PNN) can be found using the cluster's
distance derived using main-sequence (MS) fitting. The mass of the PNN can then be deduced from
theoretical core-mass\slash luminosity relations \citep[for example][]{millerbertolami16}---thus providing a point on the important initial-final mass relation \citep[e.g.,][]{Cummings2018, Barnett2021, Marigo2022}. Additionally, the chemical
composition of the PN, along with the progenitor mass, provide information about the dredge-up of processed
material from the stellar interior---theoretically predicted to depend strongly on the
initial mass of the progenitor star \citep[see the recent review by][and references therein]{Kwitter2022}.

This paper
is the second in a series in which we report membership tests for PNe that lie close to star clusters on the sky. Our tests are based on obtaining precise measurements of proper motions (PMs) of the PNNi.

In our initial study \citep[][hereafter Paper~I]{Bond_JaFu12024}, we reviewed previous attempts to confirm star-cluster membership for PNe. The criteria are that the PN has to lie within the tidal radius of the cluster, and it must have a radial velocity, interstellar extinction, estimated distance, and PM that are in agreement with those of the potential host cluster. 

In the case of PNe belonging to Galactic globular clusters (GCs), three objects had already been shown to satisfy all of the membership criteria, including the PM tests \citep[][and references therein]{Bond2020}. In Paper~I we reported PM measurements for a fourth candidate, the PN JaFu~1 lying near the Galactic-bulge GC Palomar~6. We determined the PM of its central star through the use of multi-epoch high-resolution imaging with the {\it Hubble Space Telescope\/} (\HST).  Unfortunately, we found a significant discordance between the PM of the nucleus and the mean PM of members of Palomar~6,  ruling out cluster membership for JaFu~1.\footnote{In spite of the high and remarkably similar radial velocities of the PN and cluster.}

Turning to open clusters (OCs), we note that over a dozen PNe lie tantalizingly close to Galactic OCs on the sky. However, detailed investigations have disqualified nearly all of them as members, as reviewed in depth by \citet{Majaess2007, Majaess2014}, \citet{Frew2016}, and \citet{Davis2019}, and summarized in our Paper~I\null. 

A convincing case of a PN belonging to an OC is IPHASX J055226.2+323724
(PN\,G177.6+03.0). This PN has recently been confirmed as a member of the OC M37 (NGC~2099), based on the \Gaia\/ PM of its nucleus, its nebular radial velocity (RV), and its interstellar extinction, all of which agree with those of the cluster \citep{Griggio2022, FragkouM37_2022, WernerM372023}. 

Only two further Galactic PNe, to our knowledge, remain at present as strong candidate members of OCs to add to the case of the M37 PN\null.
These are  PHR~J1315$-$6555 (hereafter PHR\,J1315; IAU designation PN G305.3$-$03.1) in the OC AL~1, and BMP~J1613$-$5406 in NGC~6067 \citep{Frew2016, FragkouBMP2019, FragkouNGC6067_2022}.  In the Local Group, a PN that likely belongs to an OC in M31 was identified by 
\citet{Bond2015} and analyzed spectroscopically by \citet{Davis2019}.

In the present paper, we report PM measurements with \HST\/ for two stars that are candidate nuclei of PHR\,J1315. For \hbox{BMP~J1613$-$5406}, at this writing we have obtained only first-epoch \HST\/ imaging of its PNN and host cluster, so a PM determination must await a second observation in 2026.


\section{The Distant Open Cluster AL~1 \label{sec:AL1} }

\citet{Andrews1967} reported their discovery of a compact (diameter $\sim$$75''$) OC in the southern constellation Musca on a long-exposure photographic plate obtained with the Armagh-Dunsink-Harvard (ADH) Schmidt telescope at the Boyden Observatory.  The cluster was also noted independently by \citet[][no.~144 in their list]{vandenBergh1975} in an imaging survey of the southern Galactic plane using blue- and red-sensitive plates from the Curtis Schmidt telescope at Cerro Tololo. It was recognized yet again by \citet{Lauberts1982} in the European Southern Observatory\slash Uppsala Survey of the ESO(B) Atlas photographs obtained with the ESO 1-m Schmidt telescope at La~Silla, Chile; in this publication the cluster is cataloged as ESO 96-SC04. Hereafter we designate the cluster as AL~1, in recognition of the first published discovery. AL~1 lies in a crowded low-Galactic-latitude field, and is centered at Galactic coordinates $(l,b)=(305\fdg4, -3\fdg2)$.


The first ground-based CCD photometric observations of Al~1 were obtained by \citet{Phelps1994}. Based on these data, 
\citet{Janes1994} estimated the cluster's distance to be 7.57~kpc, but the color-magnitude diagram (CMD) showed heavy contamination by field stars.
CCD data obtained by \citet{Carraro1995} yielded a larger distance of $d\simeq11.8$~kpc; the authors pointed out that this makes AL~1 one of the most distant known Galactic OCs. They estimated a reddening of $E(B-V)=0.75$  and an age of 700~Myr, based on isochrone fitting in the CMD\null. Additional CCD data were obtained by \citet{Carraro2004}; they updated the cluster parameters to $E(B-V)=0.7\pm0.2$, a distance of $12\pm1$~kpc, and an age of 800~Myr.\footnote{\citet{Carraro2004} designated the cluster ESO 96-SC04. Curiously, \citet{Carraro2005} published another analysis of the cluster based on a different set of data, this time calling it AL~1, and not citing their paper from a year earlier. They obtained a lower reddening of $E(B-V)=0.34\pm0.05$ and a correspondingly larger distance of $d=16.9$~kpc, but again an age of 800~Myr. \citet{Parker2011} pointed out that increasing the reddening to $E(B-V)=0.7$, as found in other studies, would reduce the distance to 10.1~kpc.}

AL~1 appears to be mildly metal-poor relative to the Sun. Based on an analysis of the \caii\ infrared triplet lines in the spectra of two cluster members, \citet{Frinchaboy2004} reported a metallicity of $\mathrm{[Fe/H]}=-0.51\pm0.30$. 

The realization that AL~1 likely hosts a PN (see next section) sparked heightened interest in the cluster. \citet{Majaess2014} used new optical and near- and mid-infrared photometry to determine the run of interstellar extinction with distance in the direction of AL~1, deriving a distance of
$10.0\pm0.4$~kpc, and an age of 800~Myr for the cluster.

\citet[][hereafter Fragkou+]{FragkouPHR2019} obtained the first multicolor \HST\/ imaging of the PN and cluster. Using CMD isochrone fitting to \HST\/ stellar photometry of cluster member stars, they determined a reddening of $E(B-V)=0.83\pm0.05$, a distance of $12\pm0.5$~kpc, and an age of $660\pm100$~Myr for the AL~1 cluster.


Most recently, \citet{Perren2022} included AL~1 in a detailed investigation of 25 OCs with distances considered to be greater than 9~kpc,\footnote{In the \citet{Perren2022} study, the cluster is called vd Bergh-Hagen 144 or BH\,144.} selected from a search of four extensive catalogs of Galactic OCs. This study exploited recently available astrometric and photometric data from the \Gaia\/ spacecraft. In particular it became possible to use PMs to reject the numerous cluster non-members in the field of AL~1, using a Gaussian-mixture model. A Bayesian analysis of the \Gaia\/ CMD of the cluster members, based on PARSEC isochrones \citep{Bressan2012}, was then used to derive metallicites, reddenings, distances, ages, and other parameters for the clusters. The results for AL~1 included a subsolar metallicity of $\rm[Fe/H]=-0.53^{+0.05}_{-0.02}$, reddening of $E(B-V)=0.83^{+0.01}_{-0.03}$, distance of $10.06^{+0.35}_{-0.27}$~kpc, and age of $1.05^{+0.02}_{-0.07}$~Gyr. 



\section{The Planetary Nebula PHR J1315 \label{sec:PHR} }

\citet{Parker2006} used wide-angle photographs of the southern Galactic plane obtained with the Anglo-Australian Observatory UK Schmidt Telescope and a narrow-band \Ha\ filter to search for  resolved nebular objects. This survey resulted in the discovery of over 900 new true or suspected PNe, including PHR~J1315.\footnote{``PHR'' stands for the surnames of the authors Q.~Parker, M.~Hartley, and D.~Russeil of the \citet{Parker2006} paper.}

\citet{Parker2011} pointed out that
PHR~J1315 lies within $23''$ of the center of AL~1. This is well within the tidal radius of the cluster, estimated by \citet{Perren2022} to be $2.8^{+0.6}_{-0.5}$~arcmin, making the PN a strong candidate for membership in the cluster. Spectra of the nebula presented by \citet{Parker2011} confirmed its PN nature. Further support for membership came from the close agreement of the radial velocity (RV) of the PN with the RVs of three bright cluster member stars; from the agreement of the distance of the PN estimated from an \Ha\ surface-brightness versus linear radius relation and the distance of the cluster (see Section~\ref{sec:AL1}); and from agreement between the reddening of the PN (estimated from the spectroscopic Balmer decrement) and the reddening of the cluster.


The remaining membership test is to verify that the PM of the central star of PHR~J1315 is consistent with the PMs of cluster member stars. For this purpose we used \HST\/ to obtain new high-resolution images of the PN and cluster, which we combined with archival images in order to make the astrometric measurements.

\section{\emph{Hubble Space Telescope\/} Observations \label{sec:HST} }

\HST\/ images of PHR~J1315 and AL~1 have been obtained at two epochs separated by 12~years, as listed in Table~\ref{tab:exposures}. All of the data are from exposures with the UVIS (ultraviolet and visible light) channel of the
Wide Field Camera~3 (WFC3), providing an image scale of 40\,mas pixel$^{-1}$. WFC3 places two adjacent 2k$\times$4k CCD chips in the focal plane, with a small gap between them. First-epoch observations were obtained in 2012, as described by Fragkou+. The full-frame two-chip ``UVIS'' aperture was used in four broad-band filters (F200LP, F350LP, F555W, and F814W), providing a $160''\times160''$ field of view centered on the AL~1 cluster. A two-point dither pointing pattern was used, along with two additional short exposures in F555W and F814W\null. In order to obtain a high-resolution image of the PN itself in the [\oiii] $\lambda$5007 F502N bandpass, the ``UVIS1'' aperture, was used, placing the nebula in the center of CCD Chip~1. A two-point dither pattern was again used.

For our second-epoch imaging in 2024, our aim was to obtain data allowing us to determine a precise PM for the central star of PHR~J1315, along with PMs of neighboring members of the cluster. We chose to image only in F555W, 
because it provides the best stellar detection efficiency among the filters used in the epoch-1 data, and there is extensive knowledge of its astrometric properties.
To maximize observing efficiency we placed the PN at the center of the ``UVIS2-2K2C-SUB'' subarray, giving a field of view of $81''\times81''$. For good sampling of the 
pixel-phase space,
we used a five-point dither pattern. Our frames were post-flashed to a level of 40 electrons per pixel, to help mitigate the effects of degraded charge-transfer efficiency (CTE) in the aging WFC3 detectors.\footnote{Cosmic-ray damage produces an accumulation of traps in the CCD detector's silicon lattice that impede charge transfer during readout.
Post-flashing the detectors fills most of these traps before the readout begins. For more details, see the WFC3 Instrument Handbook at \url{https://hst-docs.stsci.edu/wfc3ihb/chapter-6-uvis-imaging-with-wfc3/6-9-charge-transfer-efficiency}.} The epoch-1 exposures did not use post-flashing, but the background in those images is high enough that it was unnecessary.



\begin{deluxetable*}{lllcl}
\tablecaption{Log of {\it Hubble Space Telescope\/} WFC3 Observations of PHR\,J1315 and AL\,1 \label{tab:exposures} }
\tablehead{
\colhead{Date}
&\colhead{Aperture}
&\colhead{Filter}
&\colhead{Exposures}
&\colhead{Program ID}\\
\colhead{}
&\colhead{}
&\colhead{}
&\colhead{[s]}
&\colhead{and PI}
}
\startdata
2012 Mar 6  & UVIS  & F200LP & $2\times510$  & GO-12518 (Q. Parker) \\
            & UVIS  & F350LP & $2\times550$  &  \\
            & UVIS1 & F502N  & $2\times1000$  &  \\
            & UVIS  & F555W  & $2\times510$, 36  &  \\
            & UVIS  & F814W  & $2\times510$, 40  &  \\
\smallskip            
2024 Feb 11 & UVIS2-2K2C-SUB  & F555W & $5\times465$  & GO-17531 (H.E.B.) \\
\enddata
\tablecomments{All of the frames may be downloaded from the Mikulski Archive for Space Telescopes at \url{https://archive.stsci.edu}. The WFC3 frames analyzed in this paper are available at \dataset[http://dx.doi.org/10.17909/vxvg-q276]{http://dx.doi.org/10.17909/vxvg-q276}.}
\end{deluxetable*}

\section{Identification of the Central Star \label{sec:central_star} }

\begin{figure*}
\centering
\includegraphics[width=0.85\linewidth]{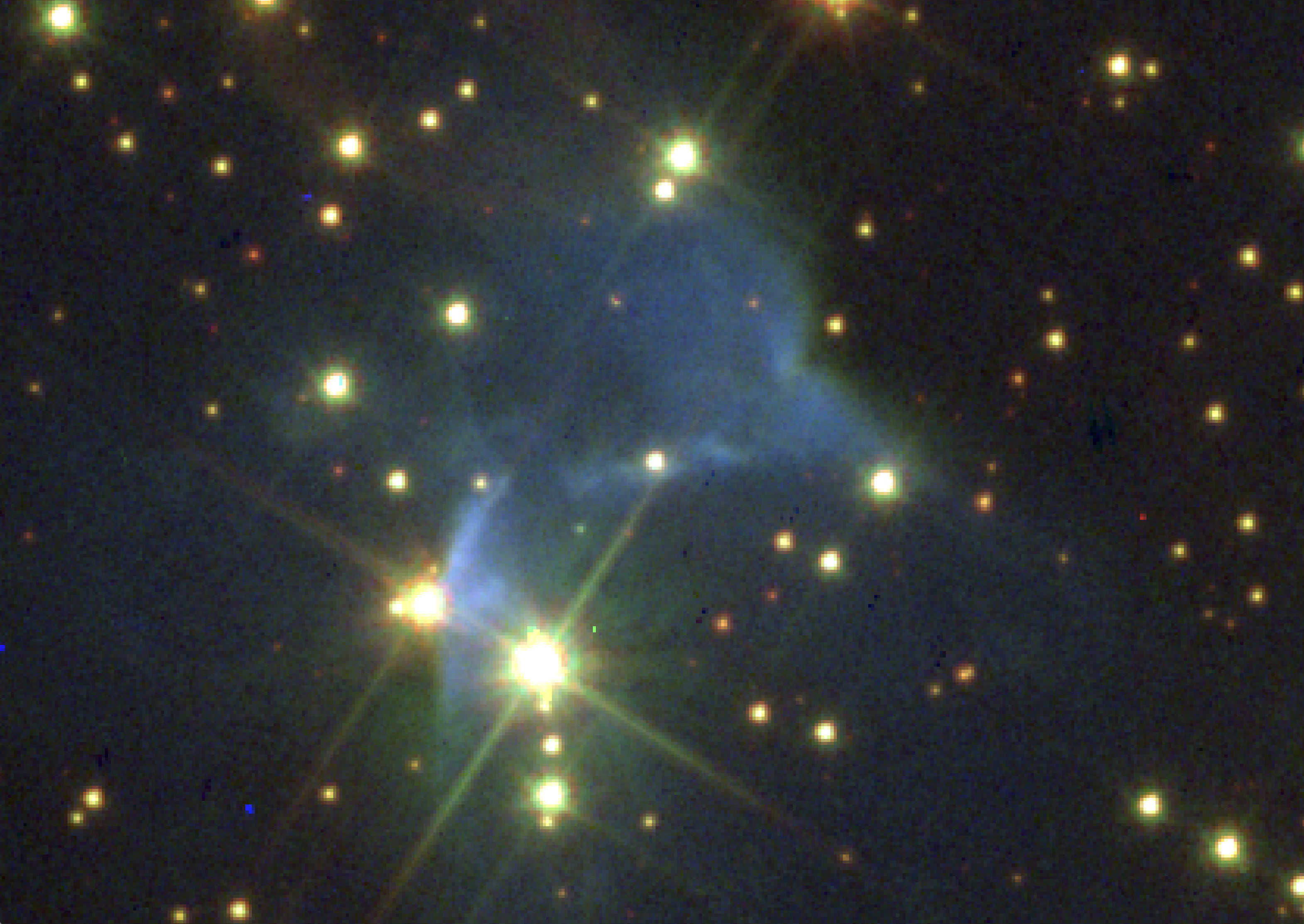}
\caption{
False-color \HST\/ image of the planetary nebula PHR J1315$-$6555. Created from WFC3 frames in F814W ($I$; red), F555W ($V$; green), and F502N ([\oiii] $\lambda$5007; blue).
North is at the top, east on the left. Height of frame is $12''$.
\label{fig:HST_PHRJ1315} }
\end{figure*}

Figure~\ref{fig:HST_PHRJ1315} presents a false-color image of PHR~J1315, which we constructed from the broad-band F814W and F555W frames (red and green, respectively) and the narrow-band F502N [\oiii] images (blue). Several alternative renditions of the \HST\/ imagery were presented by Fragkou+; here we have incorporated our new data, which improve the signal-to-noise ratio in the F555W bandpass. Note that the F555W filter does have some sensitivity to several PN emission lines, including [\oiii] $\lambda\lambda$4959,5007, \Hb, and \Ha. 

As discussed by Fragkou+, PHR~J1315 shows a bipolar morphology, with a pinched ``waist'' or equatorial ring extending from the east-southeast to the west-northwest. Two extended hollow lobes lie along an axis perpendicular to the waist, opening up with increasing distance from the center. This morphology resembles that of numerous well-known bipolar PNe, such as the remarkably similar NGC~2346.\footnote{\HST\/ imagery of NGC~2346 from the Hubble Heritage Project is available at \url{https://esahubble.org/images/opo9935d/}}

\begin{figure}
\centering
\includegraphics[width=\linewidth]{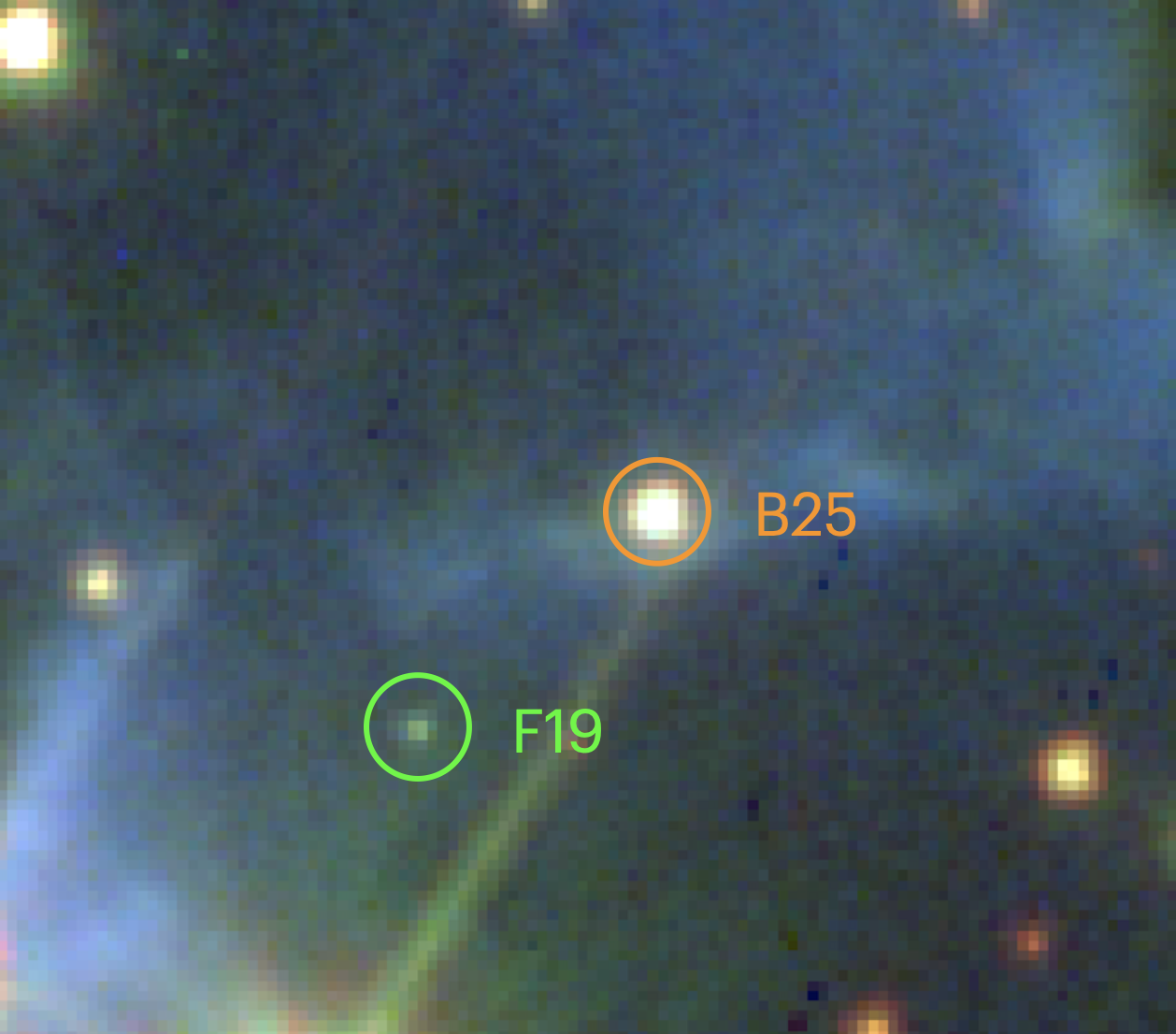}
\caption{
Central region of the nebula, enlarged from Figure~\ref{fig:HST_PHRJ1315}. Two candidate central stars are marked: ``F19'' from \citet{FragkouPHR2019}, and our proposed nucleus, labeled ``B25.'' Height of frame $4\farcs35$.
\label{fig:HST_PHRJ1315_zoom} }
\end{figure}

In order to identify the central star of PHR~J1315, Fragkou+ assumed that the PNN is extremely hot and would be very blue relative to other stars in the \HST\/ frames. Using the $m_{\rm F200LP}-m_{\rm F350LP}$ color index as a criterion, Fragkou+ selected six blue stars in the vicinity of the PN\null. One of them fell close to the center of PHR\,J1315, and Fragkou+ proposed it as the nucleus. Figure~\ref{fig:HST_PHRJ1315_zoom} zooms in on the center of the PN in our \HST\/ image. Their candidate PNN is labeled ``F19'' in the figure, and we use this designation hereafter.\footnote{\citet{Gonxzalez2021}, based on a catalog search aimed at identifying central stars of PNe using \Gaia\/ data, proposed a 15th-mag star as the PNN of PHR~J1315. Their candidate is the brightest star in our Figure~\ref{fig:HST_PHRJ1315}, lying southeast of B25. This star is located well away from the center of the PN, and it has a \Gaia-based parallax distance of only $\sim$1350~pc. We do not consider it a plausible nucleus.}

\begin{figure*}
\centering
\includegraphics[width=0.8\linewidth]{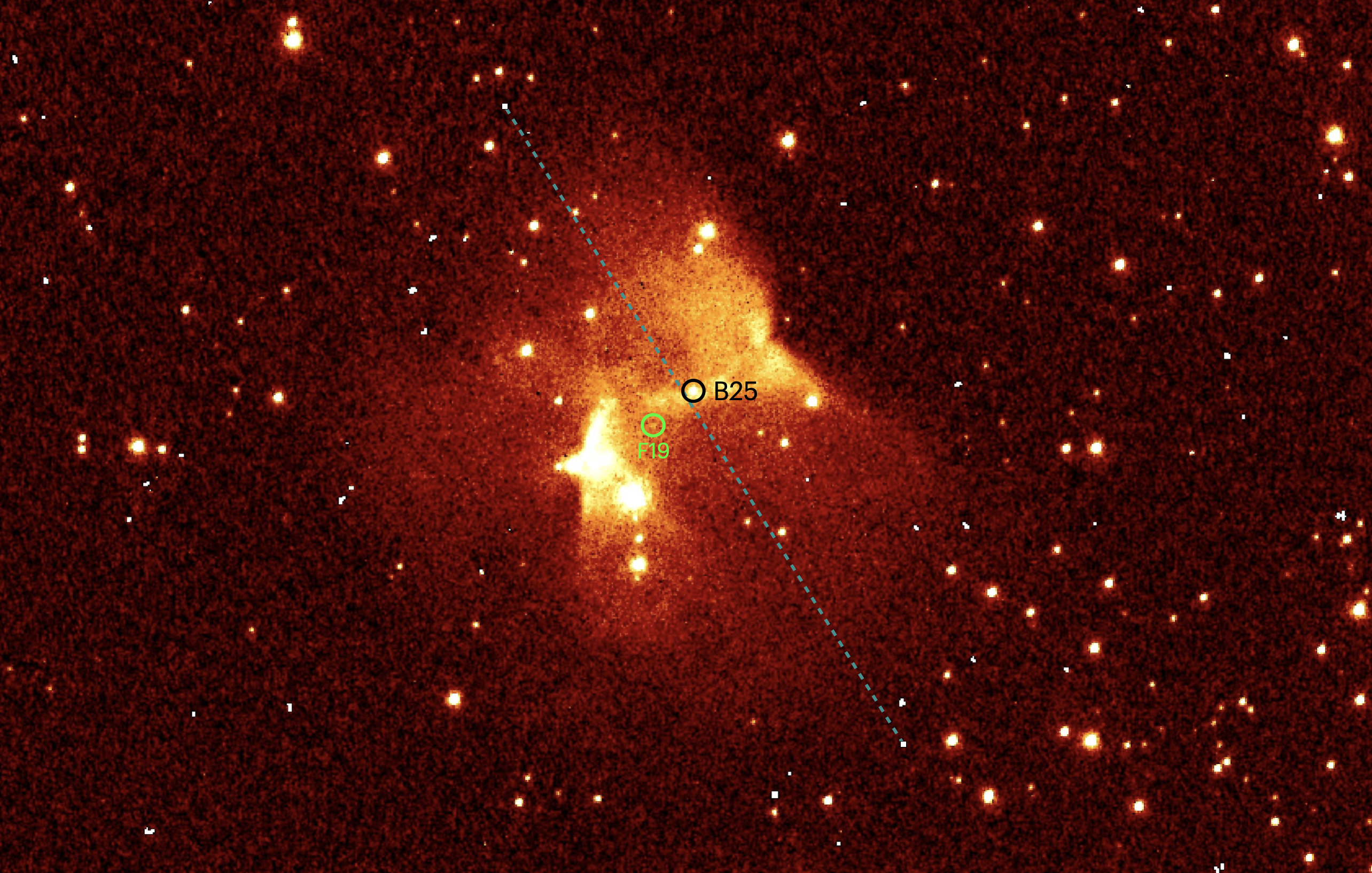}
\caption{
Heavily stretched rendition of the [\oiii] $\lambda$5007 WFC3 image. Both candidate central stars from Figure~\ref{fig:HST_PHRJ1315_zoom} are marked. A dashed cyan line lies along the approximate axis of the bipolar nebula, perpendicular to its narrow waist. Our candidate central star B25 is closer to this axis than is the F19 candidate.
\label{fig:Deep_image} }
\end{figure*}

F19 lies conspicuously off-center in the nebula, leading us to question whether it is the correct central star. In contrast, a brighter star lies close to the geometric center of the PN, and in fact it is precisely located on a strand of nebulosity extending across the nebular waist.  We designate this star ``B25,'' and it is also marked in Figure~\ref{fig:HST_PHRJ1315_zoom}. In Figure~\ref{fig:Deep_image} we show a heavily stretched rendition of the [\oiii] $\lambda$5007 image in order to illustrate the locations of B25 and F19 with respect to the axis of the PN lobes. We have drawn by eye a dashed cyan line to mark the axis in Figure~\ref{fig:Deep_image}, showing that B25 is much closer to it than F19.

B25 is at most slightly blue (as discussed below), but this does not necessarily rule it out as the true PNN\null. For example, the central star of NGC\,2346 (V651~Mon) has an A5\,V spectral type, as first shown by \citet{Mendez1978}. This star would be incapable of photoionizing the PN\null. Numerous subsequent studies \citep[e.g.,][and references therein]{Brown2019} have shown that the nucleus of NGC\,2346 is a binary system with a 16-day orbital period. It is likely a post--common-envelope binary, consisting of the A5\,V star and an optically faint but very hot companion. Another possibility is that B25 could be a wider binary (but still unresolved in the \HST\/ images) containing a MS star and a PNN, which never interacted. An example of such a system is provided by NGC~1514, whose nucleus combines an A-type star with a hot subdwarf in a binary with an orbital period of about 3300~days \citep{Jones2017}.

We list details of the two PNN candidates in Table~\ref{tab:candidates}.
B25, our preferred candidate nucleus, is contained in \Gaia\/ Data Release~3 (DR3; \citealt{Gaia2016, Gaia2023}); however, the uncertainties in its \Gaia\/ PM ($\pm$0.67 and $\pm$0.83 mas\,yr$^{-1}$ in right ascension and declination, respectively) are too large to be useful discriminants of cluster membership. The F19 candidate nucleus is too faint for inclusion in DR3.  



\begin{deluxetable}{lcc}
\tablecaption{Parameters for Candidate Nuclei of PHR\,J1315 \label{tab:candidates} }
\tablehead{
\colhead{Parameter}
&\colhead{F19}
&\colhead{B25}
}
\decimals
\startdata
RA (J2000) & 13 15 18.748 & 13 15 18.591\\
Decl (J2000) & $-65$ 55 01.07 & $-65$ 55 00.21 \\
%
$\mu_\alpha\cos\delta$ [mas\,yr$^{-1}$] & $-5.087 \pm 0.155$ & $-5.144 \pm 0.034$\\
$\mu_\delta$ [mas\,yr$^{-1}$] & $-0.256 \pm  0.093$ & $-0.365 \pm  0.027$ \\
$G$ magnitude & $\dots$ & $20.126\pm0.008$ \\
F200LP magnitude & $24.253 \pm  0.016$ & $20.717 \pm  0.008$ \\
F350LP magnitude & $24.095 \pm  0.016$ & $20.414 \pm  0.006$ \\
F555W magnitude  & $24.487 \pm  0.017$ & $20.776 \pm  0.002$ \\
F814W magnitude  & $24.032 \pm  0.060$ & $19.550 \pm  0.010$ \\
$m_{\rm F555W}-m_{\rm F814W}$ [mag] & $0.455 \pm 0.062$ & $1.226 \pm 0.010$ \\
\enddata
\tablecomments{Positions are for equinox J2000, epoch 2016.0. The $G$ magnitude for B25 is from \Gaia\/ DR3. Other quantities are from the analysis presented below. Quoted PM and magnitude errors are internal only. Magnitudes are on the Vega scale.}
\end{deluxetable}






\section{Data Analysis \label{sec:dataanalysis} }

\subsection{Astrometry \label{subsec:astrometry} }

Our astrometric data reduction and analysis steps closely followed those detailed in Paper~I (see also \citealt{Bond2020}). Briefly, source positions and instrumental magnitudes of neighbor-subtracted sources are measured in each pipeline-calibrated, unresampled WFC3 frame (filename suffix \texttt{\_flc}; these frames include corrections for detector CTE). Positions at both observation epochs are determined via effective-point-spread-function (PSF) fitting, using focus-diverse PSF models that account for telescope breathing (\citealt{Anderson18}). We used a combination of first- and second-pass reduction techniques, based on the software tools \texttt{hst1pass}\footnote{{\tt hst1pass} is available at \url{https://www.stsci.edu/hst/instrumentation/wfc3/software-tools}.} \citep{Anderson22a} and \texttt{KS2} (see \citealt{Bellini17}). Positions are corrected for the effects of geometric distortion, using the high-precision solution for WFC3\slash UVIS frames presented by \citet{Bellini11}. 
We note that not all UVIS filter bandpasses are equally suitable for precise astrometry, either because of a lack of appropriate high-quality distortion solutions, or because of chromatic effects in the shorter-wavelength filters (see discussion in Section~3 of \citealt{Bellini11}). For the data set at hand, we used only the F555W and F814W frames for the astrometric measurements.

Next, we defined a common reference frame based on stars in the WFC3 images that are also cataloged in \textit{Gaia} DR3. Here source positions are transformed using general, six-parameter linear transformations. These transformed positions are used for cross-identification of the same sources across different images.

Relative PMs are then iteratively computed as follows, using the prescriptions given in \citet{Bellini14, Bellini18a}. The locally transformed reference-frame positions of each source are fit as a function of time, using a least-squares linear relation; the slope of the relation is a direct measurement of the source's PM\null. Local transformations are based on a network of likely cluster members; this helps mitigate systematic position residuals that are mostly due to uncorrected distortion and PSF residuals and---to some extent---CTE correction residuals. Members of AL\,1 are initially selected on the basis of their locations in the CMD, with the selection improved later using PMs from the previous iterations.  

Finally, we applied {\it a-posteriori} local adjustments to the PMs, this time primarily targeting uncorrected CTE residuals. These systematic residuals are a function of local background, distance from the readout amplifier in the raw exposures, and source brightness. For each star in the PM catalog, the correction is computed as the average offset in $(\mu_\alpha \cos\delta,\ \mu_\delta)$ of the closest 15 (within a radius of 250 pixels) cluster stars with brightnesses within $\pm$1~mag of the star under consideration (see Section~7.4 of \citealt{Bellini14} for a detailed description of the procedure).\footnote{These corrections were small. The median is zero (by definition), and the semi-interquartiles for corrections in $\mu_\alpha\cos\delta$ and $\mu_\delta$ were 0.006 and $0.009\,\masyr$, respectively.} 



{

Because transformations are based on cluster members, the resulting PMs are relative to the bulk motion of the cluster, which by definition lies at the origin of the relative vector-point diagram. To convert them into absolute PMs, we applied a correction based on 110 relatively well-measured\footnote{Defined as renormalized unit weight error (RUWE) $<$ 1.4 and error in total PM $<$ $0.3\, \masyr$.} \Gaia\/ DR3 stars in common with our \HST\/ measurements. The weighted absolute PM zero-points computed this way are 
$(\mu_{\alpha} \,\cos\delta, \mu_{\delta})_{0} = (-5.150 \pm 0.171$, $-0.386 \pm 0.206)$ mas\,yr$^{-1}$.
These values also define our best estimate for the absolute PM of AL~1.\footnote{Our result agrees well with that of \citet{Perren2022}:  $(\mu_{\alpha} \,\cos\delta, \mu_{\delta})_{0} = (-5.14 \pm 0.33$, $-0.47 \pm
0.32)$ mas\,yr$^{-1}$.} The third and fourth rows in Table~\ref{tab:candidates} list the absolute PMs for the two candidate nuclei. 

\subsection{Photometry}

\subsubsection{Calibration and Errors}

We calibrated the instrumental magnitudes to the Vega-mag \HST\/ flight system, following the prescriptions given in \citet{Bellini17}. To estimate uncertainties, we generally compute internal magnitude errors as the RMS of $N$ repeated observations around the mean values, divided by the square root of $N$. For the data at hand, this can be reasonably done only for the F555W data of epoch 2. To estimate photometric errors in the other bands, we note that photometric errors scale with the S/N of sources in a very similar way for a wide variety of filters (see, e.g., the discussion in Section~5.2 of \citealt{Bellini14}). Therefore, we assigned as photometric errors in the other bands the corresponding photometric RMS of epoch-2 F555W magnitudes at the same S/N level, divided by the square root of the number of available observations. Table~\ref{tab:candidates} lists our derived Vega-scale magnitudes for the F19 and B25 candidates, with internal errors derived as described here. The uncertainties in the Vega-system zero-points are 0.0127, 0.0048, 0.0028, and 0.0056~mag in F200LP, F350LP, F555W, and F814W, respectively \citep{Calamida2021}. 

\subsubsection{Variability}

The available data are too sparse to allow a thorough investigation of possible variability of either candidate PNN\null. We note, however, that the photometric RMS scatter among the five epoch-2 F555W observations for B25 and F19 is 0.004 and 0.039 mag, respectively, in line with expectations from repeated observations of non-variable sources at the same S/N level. Additionally, the magnitude differences between the average epoch-1 and epoch-2 F555W photometry of B25 and F19 are 0.001 and 0.028 mag, respectively, again in agreement with expected photometric errors. Hence, the limited data do not indicate significant variability for either F19 or B25. 


\subsubsection{Saturated Stars \label{subsubsec:saturatedstars} }

A small number of stars are saturated in all of the F555W and\slash or F814W exposures, except in the single short exposures in each filter obtained at epoch~1. Saturation in the long exposures sets in at about $m_{\rm F555W}=18.5$ and $m_{\rm F814W}=17.5$. For the saturated stars we are unable to obtain their PMs from the \HST\/ data due to the lack of second-epoch observations; however, all of them have lower-precision PMs available in \Gaia\/ DR3.  

\section{Proper-Motion Cluster Membership Tests \label{sec:membershiptest} }

In Figure~\ref{fig:vectorpoint}, Panel~(a), we plot the absolute PMs we measured for all detected stars in the \HST\/ field. 
A conspicuous clump of points represents members of the AL\,1 cluster, centered at
$(\mu_{\alpha} \,\cos\delta, \mu_{\delta}) = (-5.150, -0.386)$ mas\,yr$^{-1}$,
as reported in Section~\ref{subsec:astrometry}.
Panel~(b) in the figure shows a zoomed-in view around the cluster members. In both panels the distribution of AL~1 stars appears anisotropic, being elongated along the declination direction (which is also almost exactly the direction of Galactic latitude). An elongation in a similar direction is also seen in the \Gaia\/ PM data for the cluster members selected by \citet{Perren2022}, in spite of the \Gaia\/ uncertainties being considerably larger than for the \HST\/ measurements. For the \HST\/ astrometry, the internal PM errors are around 0.03~$\masyr$ for well-exposed stars in the magnitude range $17<m_{\rm F555W}<23$, and exponentially increase for fainter stars, reaching $\sim$0.1 mas\,yr$^{-1}$ at $m_{\rm F555W}\simeq25$. By contrast, the PM uncertainties for \Gaia\/ are about 0.12~$\masyr$ at $G\simeq18.5$, and about 0.27~$\masyr$ at $G\simeq19.5$.

\begin{figure*}
\centering
\includegraphics[width=\linewidth]{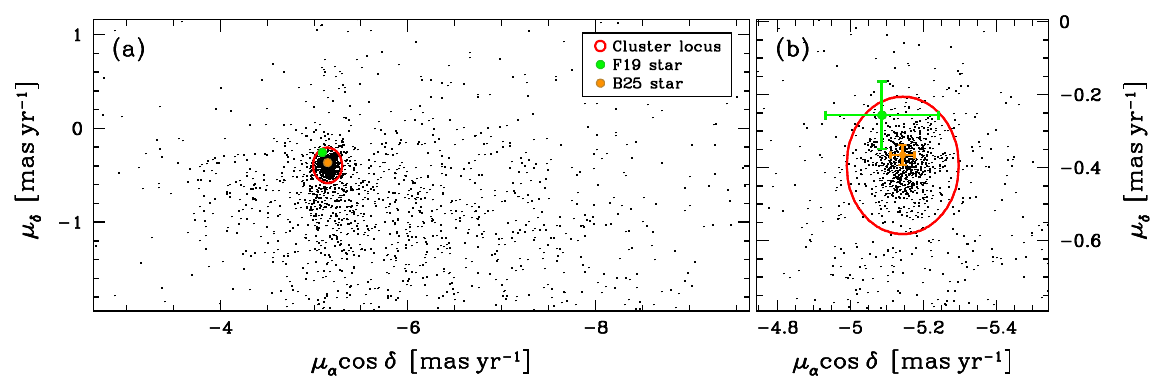}
\caption{(a) \HST-based vector-point diagram of sources in the analyzed field. The red ellipse, enclosing the compact clump of stars, is used to define AL~1 cluster membership. Its semi minor and major axes are three times the observed velocity dispersion of AL~1 stars along the R.A. and Dec.\ directions, respectively. The candidate central stars F19 (green) and B25 (orange) are marked using filled circles. 
(b) Enlarged view of the vector-point diagram around the locus of AL~1 cluster members.  Here F19 and B25 are shown with error bars. Both stars have PMs statistically consistent with cluster membership. See text for details.
\label{fig:vectorpoint} }
\end{figure*}

We define as cluster members those stars whose PMs lie within the red ellipses in both panels of Figure~\ref{fig:vectorpoint}; their semi-major and semi-minor axes are 0.240 and 0.189 mas\,yr$^{-1}$, respectively, equal to three times the observed dispersions along declination and right ascension.
In both figure panels, we highlight the locations of F19 and B25, plotted with filled circles. Error bars for the two candidates are shown only in Panel~(b).

Figure~\ref{fig:vectorpoint} confirms that our favored central star, B25, has a PM fully consistent with cluster membership; in fact, it is situated close to the center of the cluster distribution. The PM of F19 also agrees within its larger uncertainties with the locus of cluster stars, but less conclusively than for B25.

\section{Properties of the AL\,1 Cluster}

\subsection{Correcting for Differential Reddening}

In this section we examine the CMD of AL\,1, and infer from it the age and other parameters of the cluster. The CMD is based on our \HST\/ photometry, and is for stars selected to be cluster members based on their PMs detertmined as described in the previous section. 

First, however, we need to take spatially variable interstellar reddening into account. As noted at the end of Section~\ref{sec:AL1}, the average reddening for AL~1, based on \Gaia\/ photometry, is about $E(B-V)=0.83$. However, the reddening is clearly quite patchy, as shown by examination of direct images,\footnote{AL~1 lies within about $2^\circ$ of the southeast edge of the Coalsack Nebula.} and by an observed broadening of the CMD of cluster members along the direction of the reddening vector. To mitigate this effect, we applied a differential-reddening (DR) correction to the stars in the WFC3 field. We followed the prescription given in \citet{Milone12}; see also \citet{Bellini2017drc} for a detailed description of the method.

The correction procedure is based on the assumptions that the cluster stars are all at the same distance from us, and that the variable extinction lies between the cluster and us (i.e., it is not located within the cluster). 
The method is based on a local CMD correction along the reddening vector, using a network of well-measured cluster stars (called the reference stars); and it only removes the differential part of the reddening over the field with respect to the average reddening of the reference stars. It does not remove the entire reddening.


The DR reference stars should ideally lie on the CMD sequence where it is perpendicular to the reddening direction, which is where the effects of DR on the broadening of a CMD sequence are maximized. The best such reference stars lie in the vicinity of, and just above, the main-sequence turnoff (MSTO). Unfortunately, however, in our data the brighter stars around the MSTO are saturated in all of our frames, except for the two short exposures obtained only at epoch~1, as discussed in Section~\ref{subsubsec:saturatedstars}. Thus these stars are missing from our PM catalog, which we use to select cluster members. We therefore had to rely on a less optimal set of fainter reference stars, from the bright end of the MS just below saturation, down to $m_{\rm F814W}\simeq23$. Cluster members are defined as those lying within the red ellipse in Figure~\ref{fig:vectorpoint}, and within a generous color band in the $m_{\rm F814W}$ versus $m_{\rm F555W}-m_{\rm F814W}$ CMD centered on the MS of AL~1. 



The choice of the number of reference stars to be used for the correction is a trade-off between having adequate statistics, but a correction able to account for small-scale spatial variations of the reddening. In our case, we are limited by the available number of selected {\it bona-fide\/} members (555) in our relatively small second-epoch field ($\approx$$1\farcm3 \times 1\farcm3$). Typically, DR corrections in the literature are based on 50--100 reference stars per spatial element, but in our case we had to rely on generally smaller samples of about 15 to obtain the best results (i.e., a CMD where AL~1 stars have the narrowest MS).

The uncorrected CMD is then rotated so as to make the reddening vector horizontal.\footnote{We adopt $A_{\rm F555W}/E(B-V)=3.227$ and $A_{\rm F814W}/E(B-V)=1.855$, taken from the Padova theoretical-isochrones website referenced below in Section~\ref{sec:isochrone_fitting}.}
This simplifies the correction procedures, since in this rotated plane DR effects produce a local horizontal shift in stellar ``colors.'' To further simplify the procedure, we fitted a fiducial curve to selected reference stars in the rotated CMD, and the color of the fiducial is then subtracted from each star. In this resulting rotated and rectified plane, reference stars form a vertical sequence with average ``color'' = 0. Here we can use the local average $\delta$``color'' of reference stars at any location to infer the local DR at that location. Once the ``color'' of each star in the field is adjusted using this method, the plane is rotated back to its original 
$m_{\rm F814W}$ versus $m_{\rm F555W}-m_{\rm F814W}$ form, which now shows DR-corrected magnitudes and colors.\footnote{Here we work in the $m_{\rm F814W}$ versus $m_{\rm F555W}-m_{\rm F814W}$ plane, rather than the more conventional $m_{\rm F555W}$ versus $m_{\rm F555W}-m_{\rm F814W}$, because the slope of the reddening vector is shallower in the former.} Note that the DR corrections are applied to all stars in the field, including the nonmembers, even though some of them could be in front of some of the extinction.

\begin{figure*}
\centering
\includegraphics[width=\linewidth]{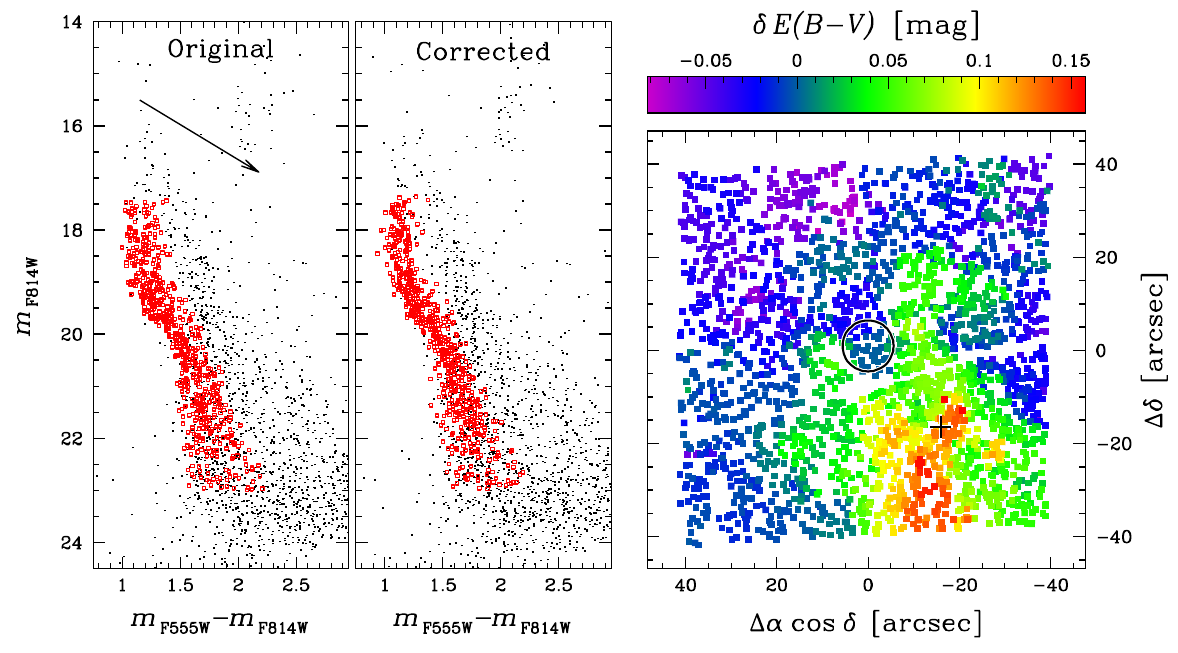}
\caption{\textit{Left:} Observed $m_{\rm F814W}$ versus $m_{\rm F555W}-m_{\rm F814W}$ CMD of stars in the \HST\/ field. The chosen proper-motion-selected reference cluster members are highlighted in red, and the black points are stars considered to be non-members. An arrow shows the direction of the reddening vector. 
\textit{Center:} DR-corrected CMD\null. The main sequence of AL~1 is now narrower than in the uncorrected CMD\null. 
\textit{Right:} Spatial map of the DR corrections, centered on the B25 candidate PN nucleus. Each star is color-coded according to the shift of nearby reference stars along the reddening vector direction with respect to the average reddening over the whole field. The color bar at the top indicates the color-coding of the points, giving the local value of $E(B-V)$ relative to the mean $E(B-V)$ across the field. A circle of radius $5\farcs6$ centered on B25 indicates the location of the PN\null. The plus sign to its southwest marks the center of the AL~1 cluster. See text for details and discussion.
\label{fig:DRC} }
\end{figure*}

The left panel of Figure~\ref{fig:DRC} shows the uncorrected $m_{\rm F814W}$ versus $m_{\rm F555W}-m_{\rm F814W}$ CMD of all stars in the \HST\/ field. The selected reference cluster members are shown as open red squares. All other sources are in black. Our PM catalog does not contain the saturated sources, so we added them back into this CMD using measurements from the first-epoch short-exposure photometry. The direction of the reddening vector is shown, for reference. The corresponding DR-corrected CMD is plotted in the middle panel.  Note that the MS of AL~1 members, in red, is now narrower than in the uncorrected CMD, demonstrating that the DR correction procedure has improved the photometric quality of the cluster CMD. 



The right-hand panel in Figure~\ref{fig:DRC} plots a spatial map of the 
inferred DRs.
Here the coordinate system has its origin at star B25. Each star has been color-coded according to the
average $\delta E(B-V)$ of neighboring reference stars (see color bar at top).
A circle of diameter $11\farcs2$ (the angular diameter of PHR\,J1315 given in the HASH catalog\footnote{The Hong-Kong/AAO/Strasbourg/H$\alpha$ Planetary Nebulae Database at \url{http://hashpn.space}; \citet{Parker2016}.}), centered on B25, is used to indicate the location of the PN\null. A black cross marks the center of the AL~1 cluster. The diagonal gap south of the PN is due to the spacing between the two UVIS chips in the epoch-1 observations. 


The map in Figure~\ref{fig:DRC} confirms the patchy nature of the extinction in front of AL\,1. The extinction is relatively small at the site of the PN PHR~J1315, but unfortunately increases toward the southwest and is highest around the center of the cluster and further southward.

Although the corrected CMD does show an improvement with respect to the uncorrected one, the correction itself is based on just $\sim$15 local reference neighbors of each star, resulting in a typical per-star correction error of 
about 0.025--0.030 mag for both the F555W and F814W magnitudes.

\subsection{Isochrone Fitting and Cluster Parameters \label{sec:isochrone_fitting} }

Figure~\ref{fig:cmd+isochrone} plots the CMD, $m_{\rm F555W}$ versus $m_{\rm F555W}-m_{\rm F814W}$, for stars that we consider to be probable members of AL\,1. Red points represent objects whose \HST-based PMs place them inside the red ellipses in Figure~\ref{fig:vectorpoint}. The blue points are for bright stars that are saturated in all but the short WFC3 exposures obtained at epoch~1, as discussed in Section~\ref{subsubsec:saturatedstars}; for these stars we do not have \HST\/ PMs, so instead we required their PMs cataloged in \Gaia\/ DR3 to fall within the ellipses of Figure~\ref{fig:vectorpoint}. All of the photometry has been corrected for DR, as described in the previous subsection.  

Data for the two candidate central stars of PHR\,J1315 are plotted in Figure~\ref{fig:cmd+isochrone} using filled circles, color-coded as indicated in the figure's legend. Uncertainties in these points are smaller than the plotting symbols, except for the color of F19. In this plot the observed magnitudes listed in Table~\ref{tab:candidates} for these two stars, which are spatially close together, have been adjusted brighter by 0.066~mag in $m_{\rm F555W}$ and 0.038~mag in $m_{\rm F814W}$, in order to correct for DR\null. 



\begin{figure}
\centering
\includegraphics[width=\linewidth]{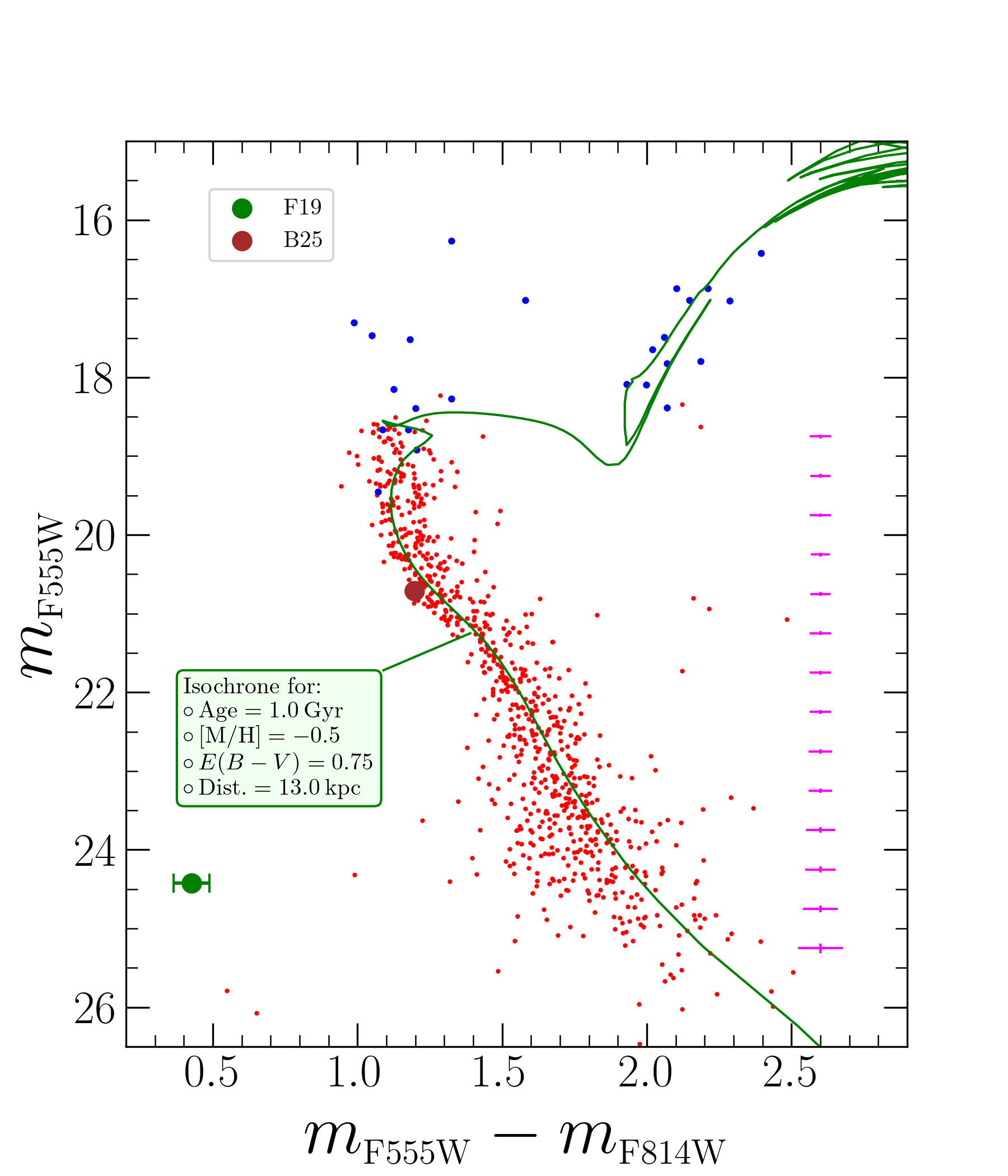}
\caption{
CMD for members of the cluster AL 1 in the \HST/WFC3 F555W and F814W bandpasses.  Red points are stars whose \HST-based proper motions place them inside the ellipses in Figure~\ref{fig:vectorpoint}; blue points are bright stars that are saturated in all but two short epoch-1 exposures, also selected to have \Gaia\/ DR3 proper motions lying inside the Figure~\ref{fig:vectorpoint} ellipses. Filled circles plot the photometry for the two candidate nuclei of PHR\,J1315, color-coded as indicated in the legend at the top. All photometry has been corrected for differential reddening, as discussed in the text. Magenta error bars show the mean uncertainties in magnitude and color, including uncertainties in the differential-reddening corrections, as a function of magnitude. The green line is a theoretical isochrone calculated for parameters given in its legend.
\label{fig:cmd+isochrone} }
\end{figure}

The cluster's CMD shows a well-populated MS and MSTO\null. The bright stars include several apparent blue stragglers and two ``yellow'' stragglers (or conceivably field stars that coincidentally have PMs similar to that of the cluster). The subgiant branch contains about a dozen evolved objects.

We fitted the CMD of AL\,1 with theoretical isochrones, obtained using the online ``PARSEC'' tool\footnote{PAdova and tRieste Stellar Evolution Code: CMD version 3.8, available at the website \url{http://stev.oapd.inaf.it/cmd}} \citep{Bressan2012, Chen2019}. This facility provides isochrones in the WFC3 Vega-based photometric system, corrected for a given amount of interstellar extinction. The PARSEC code assumes a value of $R_V=3.1$, and applies reddening calculated for the energy distribution of each star in the isochrone. We fixed the stellar metallicity at $\rm[M/H]=-0.5$, based on spectroscopy of a small number of bright cluster members (see Section~\ref{sec:AL1}). We then adopted preliminary values for the cluster's age, reddening, and distance from the studies discussed in Section~\ref{sec:AL1}, emphasizing the more recent ones. Finally we made small adjustments to these parameters in order to obtain the best fit, which is plotted as a green line superposed on the CMD in Figure~\ref{fig:cmd+isochrone}. This isochrone has an age of 1.0~Gyr, a distance of 13.0~kpc, and a reddening of $E(B-V)=0.75$. For these parameters, the initial mass of stars leaving the AGB is $2.1\,M_\odot$. We stress that our adopted parameters should be regarded with caution because of the large and variable amount of extinction, and because the spectroscopic metallicity of the cluster is uncertain at a level of about $\pm$0.3~dex (Section~\ref{sec:AL1}). 

Note that the amount of reddening given above is the mean value for stars in the \HST\/ field; at a given field location the differential amounts of reddening, mapped in Figure~\ref{fig:DRC}, should be added in order to obtain the total reddening at that site. For example, the total reddening at the center of the cluster is higher than the mean value by about 0.1~mag, i.e., it is about $E(B-V)=0.85$. This is in reasonable agreement with several independent estimates of the reddening determined from the cluster stars, as summarized in Section~\ref{sec:AL1}, considering that these studies would have preferentially used stars close to the center. The right panel in Figure~\ref{fig:DRC} indicates that the reddening at the site of the PN is close to the field mean of $E(B-V)=0.75$. \citet{Parker2011} made a completely independent estimate of the reddening of the PN based on the Balmer decrement in its spectrum, obtaining $E(B-V)=0.83\pm0.08$, in statistical agreement with our result.

As other authors have already noted, AL\,1 is one of the most distant known Galactic OCs.
At a distance from us of 13.0~kpc, the distance of AL\,1 from the Galactic center is about 10.6~kpc, putting it well outside the solar circle; and the cluster lies about 700~pc below the Galactic plane. These values lead us to anticipate that the cluster would have a moderately sub-solar metallicity.

\subsection{The Candidate Central Stars}

As discussed in Section~\ref{sec:membershiptest}, both of the possible central stars of PHR\,J1315 have PMs consistent with cluster membership. In the CMD in Figure~\ref{fig:cmd+isochrone} our candidate PNN, B25, lies at the blue edge of the cluster's MS\null. This may suggest that it has a faint and hot binary companion---making it if so an analog of the nucleus of NGC~2346, discussed above in Section~\ref{sec:central_star}. B25 likewise lies at the blue edge of the MS if we plot (not shown) $m_{\rm F200LP}-m_{\rm F555W}$ versus $m_{\rm F555W}-m_{\rm F814W}$. Its absolute magnitude of $M_{\rm F555W}\simeq+2.9$ corresponds to that of an early F-type dwarf, with a mass of roughly $1.5\,M_\odot$. 

The alternative candidate nucleus proposed by Fragkou+, F19, lies at a position in the cluster CMD consistent with that of a hot white dwarf.

\section{Summary and Discussion}

\subsection{PHR\,J1315 is a Member of Al\,1}

We summarize our findings as follows. The PN PHR\,J1315 lies close on the sky to the center of the distant Galactic OC AL\,1. As discussed in Section~\ref{sec:PHR}, the RVs, and the interstellar reddenings, of the PN and cluster agree well with each other. \citet{Parker2011} used a statistical relation between \Ha\ surface brightness and nebular radius to estimate a distance to the PN of $10.5\pm3.4$~kpc, agreeing within its uncertainty with our independently derived cluster distance of 13.0~kpc. The remaining membership test is to show that the PM of the central star of the PN agrees with that of the cluster.

Two candidate nuclei of the PN have been identified: the hot white dwarf, F19, advocated by Fragkou+; and a MS star, B25, which we prefer because of its central location in the PN\null. We have used precise PMs, measured using high-resolution \HST/WFC3 frames obtained over a 12-year interval, to verify that both stars have PMs consistent with cluster membership (although the PM uncertainties are relatively large for the faint F19). Thus all of the available information is completely consistent with the conclusion that PHR\,J1315 is a member of the star cluster. 

If B25 is the true central star of PHR\,J1315, it is almost certainly a binary companion of a hot star responsible for photoionization of the PN\null. In fact, there is ample evidence that many PNe with bipolar morphologies arise from interacting close binaries \citep[e.g.,][and references therein]{BondLivio1990, DeMarco2009, BoffinJones2019}. Thus it is plausible that B25 could be the optically dominant star in a close post-common-envelope binary. Or, as discussed in Section~\ref{sec:central_star}, it could be a wide binary whose components never interacted, containing an F-type MS star and the hot nucleus---although this would leave the origin of the bipolar morphology unexplained. It would be somewhat puzzling if the apparently single white dwarf F19 were the nucleus. (Fragkou+ found no compelling evidence for a companion of this star.) 

PHR\,J1315 joins only two other convincing cases of a PN belonging to a Galactic OC, and three instances of a PN being a member of a Galactic GC (see our Introduction). The astrophysical importance of such objects is that the initial mass of the stellar progenitor of the PN can be determined based on the age of the cluster, deduced from isochrone fitting---which is impossible for PNe in the field. For the cluster parameters we obtained (Section~\ref{sec:isochrone_fitting}) the stars currently evolving off the AGB in AL\,1 had initial masses of $2.1\,M_\odot$. This can be compared with the initial masses of $2.8\,M_\odot$ for the PN in M37 \citep{WernerM372023}, $3.4\,M_\odot$ for the PN in an OC in M31 \citep{Davis2019}, and $5.6\,M_\odot$ for the PN in NGC\,6067 \citep{FragkouNGC6067_2022}. 

The strong [\nii] emission lines in the spectrum of PHR\,J1315 \citep{Parker2011} may be even more of a puzzle than they are in the M31 PN\null. 
This is because
nitrogen-rich (and usually bipolar) PNe belonging to the ``Type~I'' classification \citep{Peimbert1978} are generally considered to arise from fairly massive progenitor stars \citep[e.g.,][and references therein]{Torres-Peimbert1997, Davis2019, Kwitter2022}}.
Theoretical studies \citep[e.g.,][]{Karakas2018} suggest an initial mass as high as $3.75\,M_\odot$ is needed for ``hot-bottom burning'' to produce nitrogen enhancements. With an initial mass as low as $2.1\,M_\odot$, PHR\,J1315 challenges this picture.


Lastly, as just noted, the range of progenitor masses for the four PNe belonging to OCs is 2.1 to $5.6\,M_\odot$. This is strikingly discordant with statistical arguments by \citet{Badenes2015} showing that a large majority of field PNe in the Large Magellanic Cloud had initial masses of 1.0 to $1.2\,M_\odot$, with only a small fraction arising from more massive stars. A possible explanation for the higher masses that dominate the progenitors of PNe found in clusters may be that most OCs dissipate into the field on a timescale shorter than the evolutionary lifetimes of their relatively low-mass stars. Thus the PNe that are discovered in OCs will preferentially arise from the minority population of younger clusters with higher-mass progenitors. As \citet{Kwitter2022} teach us, ``a PN located in a star cluster is a rare celestial gift.''

\subsection{Future Studies}

PHR\,J1315 is an important object because PNe that are verified to be members of star clusters are extremely rare. However, its astrophysical utility is compromised by several unfortunate shortcomings. It belongs to one of the most distant known open clusters, and the foreground interstellar extinction is both large and spatially variable. And it has two candidate central stars, neither of which is conclusively established to be the true nucleus of the PN.

Several followup studies would be of value. A spectroscopic investigation of the two possible central stars might help decide which is the correct PNN, and would provide constraints on its atmospheric parameters---and a new point on the initial-final mass relation. However, such a study would be challenging because both stars are crowded and faint---especially the white dwarf F19. Ultraviolet studies aimed at identifying the nucleus would be hampered by the large amount of interstellar reddening. A more promising investigation might be to conduct photometric monitoring of our candidate PNN, B25, which could be a periodic variable if it is indeed a post-common-envelope binary. But if this is the case, the hot companion's light may be overwhelmed by that of the F star, possibly even in the ultraviolet; and the evolutionary history of the system would be more complex than for a single star. 

A more robust determination of the metallicity of the cluster AL\,1 would be extremely valuable in constraining the isochrone fitting, and thus the initial mass of the progenitor. This would be relatively easy, since there are a number of bright subgiant members of the cluster. Likewise a detailed chemical-abundance study of the PN would be useful, since it is one of the very few PNe with a known progenitor mass.



\null\smallbreak

\acknowledgments

Based on observations with the NASA/ESA {\it Hubble Space Telescope\/} obtained from the Data Archive at the Space Telescope Science Institute (STScI), which is operated by the Association of Universities for Research in Astronomy, Incorporated, under NASA contract NAS5-26555. Support for Program number GO-17531 was provided through grants from STScI under NASA contract NAS5-26555.

This work has made use of data from the European Space Agency (ESA) mission
{\it Gaia\/} (\url{https://www.cosmos.esa.int/gaia}), processed by the {\it Gaia\/}
Data Processing and Analysis Consortium (DPAC,
\url{https://www.cosmos.esa.int/web/gaia/dpac/consortium}). Funding for the DPAC
has been provided by national institutions, in particular the institutions
participating in the {\it Gaia\/} Multilateral Agreement.





\bibliography{PNNisurvey_refs}

\end{document}